# The effect of sorbed hydrogen on low temperature radial thermal expansion of single-walled carbon nanotube bundles

A. V. Dolbin[1], V. B. Esel'son[1], V. G. Gavrilko[1], V. G. Manzhelii[1], S. N. Popov[1], N. A. Vinnikov[1], B. Sundqvist[2]

[1] B. Verkin Institute for Low Temperature Physics and Engineering of the National Academy of Sciences of Ukraine, 47 Lenin Ave., Kharkov 61103, Ukraine
[2] Department of Physics, Umea University, SE - 901 87 Umea, Sweden

Electronic address: dolbin@ilt.kharkov.ua

## Abstract

The effect of a normal $H_2$ impurity upon the radial thermal expansion $\alpha_r$ of SWNT bundles has been investigated in the interval T = 2.2—27 K using the dilatometric method. It is found that $H_2$ saturation of SWNT bundles causes a shift of the temperature interval of the negative thermal expansion towards lower (as compared to pure CNTs) temperatures and a sharp increase in the magnitude of $\alpha_r$ in the whole range of temperatures investigated. The low temperature desorption of $H_2$ from a powder consisting of bundles of SWNTs, open and closed at the ends, has been investigated.

## Introduction

Since their discovery in 1991 [1], carbon nanotubes have been stimulating immense research activities in the scientific community. Much interest has been focused on the unique properties of carbon nanotubes. Nevertheless, some of them are still lacking unambiguous experimental evidence. This is true in particular for the thermal expansion of single-walled carbon nanotubes (SWNTs) below room temperature. Previously [2,3], a dilatometric measurement of the radial thermal expansion coefficients $\alpha_r$ was performed for pure and Xe-saturated SWNT bundles in a temperature interval of 2.2—120 K. It was found that the coefficients $\alpha_r$ of pure SWNT bundles were positive above 5.5 K and negative at lower temperatures. The reason may be as follows. At low temperatures the thermal expansion of SWNT bundles is dominated by the lowest-frequency vibrations of individual nanotubes, which normally occurs in two-dimensional systems and is described by a negative Grüneisen coefficient [4]. However, at higher temperatures the dominant contribution to the thermal expansion is caused by temperature-induced variations of the intertube gaps in a bundle, which is similar to the interlayer thermal expansion of graphite and is described by a positive Grüneisen coefficient [5]. Doping nanotubes with Xe caused a sharp increase in the magnitudes of $\alpha_r$ practically in the whole range of the temperatures investigated. Also, a peak was observed in the dependence $\alpha_r(T)$ in the interval 50—65 K. The increase in $\alpha_r$ can occur because the Xe impurity enhances the three-dimensional character of the system and suppresses the negative contribution to the thermal expansion from the transverse acoustic vibrations perpendicular to the nanotube surfaces. The peak in the dependence $\alpha_r(T)$ can be due to the spatial redistribution of the Xe atoms in the SWNT bundles on heating the sample [6,7].

It is interesting to investigate how the radial thermal expansion of SWNT bundles can be influenced by the hydrogen impurity whose molecules differ from the Xe atoms in size (their masses and diameters are considerably smaller) and in interaction with nanotubes [8]. The relatively small $H_2$ molecules can penetrate into areas of SWNT bundles that are almost inaccessible for Xe. The penetrating power of $H_2$ molecules and their possible chemical interaction with nanotubes [8] can increase (as compared to the case of Xe) the molar concentration of the impurity (deposit) in a SWNT bundle. The higher concentration of the deposit and the attendant change in the configuration of the areas covered with the impurity molecules in the bundle will certainly affect the thermal expansion of the system.

Many researchers have looked hopefully upon carbon nanotubes as promising sorbents of $H_2$ [8,9]. This subject has been investigated thoroughly during the latest 15 years. The interest in $H_2$—CNT systems was upheld additionally by the anomalous energies of the interaction (bond break) between CNTs and $H_2$ (20—40 kJ·mol$^{-1}$ [10]). As reported in [10], the energies of the $H_2$—CNT interaction are about an order of magnitude higher than the typical interaction energies (break of Van der Waals bonds) on physical adsorption of $H_2$ in other carbon materials [11,12] and about an order of magnitude lower that the energies of rupture of covalent C—H bonds on chemosorption [13,14].

In recent years much effort has been applied to understanding possible mechanisms of such "strong" physical (or "weak" chemical) interaction between $H_2$ and carbon nanomaterials. However, the ample experimental information accumulated during 1997—2005 about the capacity of CNTs to sorbe $H_2$ suffers significant (up to three orders of magnitude, e.g., see Surveys [8,9,15,16-18]). Therefore, the fundamental aspects (mechanisms and characteristics) of the $H_2$—CNT interaction are still topical and call for further consideration. In this context, it is expedient to investigate the influence of $H_2$ upon the properties of CNTs and their bundles at low temperatures, i.e. in the region where the effects related to the interaction between a gas impurity and CNTs in different areas of their surface become most conspicuous.

The problems mentioned have motivated the goal of this study to investigate the influence of the $H_2$ impurity upon the thermal expansion of SWNT bundles at low temperatures, examining in parallel the $H_2$ desorption from SWNT bundles.

# 1. Low temperature desorption of hydrogen impurity from a powder of carbon nanotubes

## 1.1. Experimental technique

The desorption of $H_2$ from a powder of bundles of single-walled carbon nanotubes was investigated in an interval of 15—300 K using a low-temperature vacuum desorption gas analyzer. The experimental technique and the setup are detailed elsewhere [3]. According to the manufacturer, the starting SWNT powder (CCVD method, Cheap Tubes, USA) contained over 90% of SWNTs, their average outer diameter being 1.1 nm. Two samples were investigated — the starting SWNT powder and a SWNT powder after an oxidative treatment. This was applied to open the ends of the nanotubes. A detailed description of the oxidative treatment was given earlier [3].

Bundles of closed and open nanotubes were saturated with hydrogen through the same procedure. Prior to measurement, each sample was evacuated for 72 hours directly in the measuring cell of the gas analyzer to remove possible gas impurities. Then the measuring cell with the sample was filled with $H_2$ to the pressure 20 Torr and cooled down to 12 K. The used normal hydrogen was 99.98% pure and contained $O_2 \leq 0.01\%$ and $N_2 \leq 0.01\%$ as impurities. The saturation of CNTs with $H_2$ was continued at T = 12 K. Hydrogen was fed to the measuring cell until the pressure was brought to an equilibrium value of 0.1

Torr, which was no higher than the equilibrium pressure of $H_2$ vapor at 12 K. The gas not absorbed by the SWNT powder was removed from the measuring cell and the $H_2$ desorption from the CNTs was investigated. The quantities of the desorbed gas were measured on heating the sample in steps of 5 K. The hydrogen released during this stepwise heating was taken to an evacuated calibrated vessel whose internal pressure was measured using a capacitive MKS-627B pressure transducer. The gas was withdrawn from the vessel until the pressure over the CNT sample was reduced to 0.01 Torr. Then the measurement procedure was repeated at the next temperature point.

## *1.2. Results and discussion*

A diagram describing the temperature dependence of the $H_2$ quantities released from the SWNT bundles before and after the oxidative treatment is shown in Fig. 1.

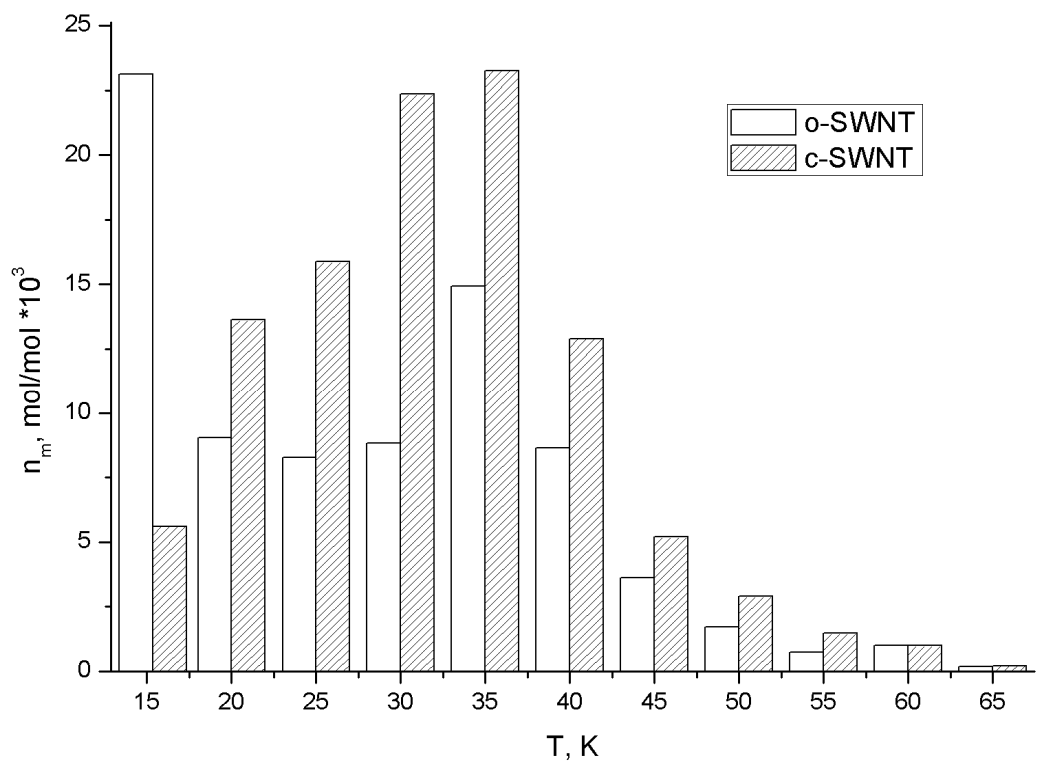

Fig. 1. Temperature distribution of $H_2$ quantities desorbed from the powder of closed (solid columns) and open (empty columns) CNTs (mole per mole of SWNT powder, i.e. the number of $H_2$ molecules per carbon atom).

The total quantity of $H_2$ desorbed from the starting and oxidated powders of SWNT bundles is given in Table 1 along with the corresponding data for Xe [3].

Table 1. The total quantities of $H_2$ and Xe impurities desorbed from our carbon nanotube sample.

| Impurity | c-SWNT | | o-SWNT | |
|---|---|---|---|---|
| | mol/mol, % | mass % | mol/mol, % | mass % |
| $H_2$ | 10 | 1.67 | 8.07 | 1.35 |
| Xe | 1.64 | 7.38 | 4.71 | 21.2 |

It should be noted that the quantity of $H_2$ desorbed from the sample and the quantity of $H_2$ sorbed by the sample on saturation were equal within the experimental error, which indicates a practically complete removal of the $H_2$ impurity from the sample.

However, the sorption reversibility can hardly be conclusive evidence for its physical origin because there is a possibility of reversible chemical sorption [8]. Nevertheless, the complete desorption of hydrogen at temperatures below 70 K is in favor of physical sorption. As was found previously [3], the air-oxidative treatment of a powder of SWNT bundles led to opening the nanotubes, which enhanced their sorptive capacity for the Xe atoms. A similar effect might be expected for $H_2$ sorption as well. However, the investigations of the sorptive capacity showed that the oxidative treatment applied to open the ends of the nanotubes did not increase the total quantity of $H_2$ sorbed by the SWNT bundles (see Table 1). This is most likely because the nanotubes of the starting powder had surface defects before the oxidative treatment (these were mainly vacancies in the carbon structure). Small $H_2$ molecules can penetrate inside nanotubes not only through their opened ends but also through the interstitial channels [16] and the vacancies in the CNT walls. Narrow interstitial channels are prohibitive [19,20] for relatively large Xe atoms which cannot therefore come inside nanotubes with closed ends.

The decreased total quantity of $H_2$ desorbed from the SWNT bundles can be due to the chemical interaction between the carbon atoms and the neighboring vacancies in the process of the oxidative treatment. The produced oxidates can block the interstitial channels [21] preventing the gas impurity molecules from occupying these sites. As a result, the total quantity of $H_2$ sorbed by the CNTs diminishes.

The $H_2$ desorption from the oxidated SWNT bundles increased sharply at the lowest temperature of desorption (15 K). According to Table 2, the $H_2$ molecules disposed at the surface of SWNT bundles have the lowest energy of binding to carbon atoms. The enhanced $H_2$ desorption at T = 15 K can therefore be attributed to a partial disintegration of the bundles in the course of oxidation [22]. As this occurs, the bundles can become smaller in size, but they increase in number. Hence, the total outer surface of the bundles also increases [23].

Table 2. The energy of impurity molecule — carbon atom binding for different dispositions of impurity particles in SWNT bundles (I — inner surface of a nanotubes; IC — interstitial channels of nanotube bundles; G — the grooves; S — surface of a bundles).

| | Binding energy | | | |
|---|---|---|---|---|
| | **I** | **IC** | **G** | **S** |
| **$H_2$** | 6 kJ/mol [16] | 11.5 kJ/mol[16] | 8.6 kJ/mol [16] | 4.7 kJ/mol [16] |
| **Xe** | 24.0 kJ/mol [19] 26.8±0.6 kJ/mol [21] | <0 [19] | 23.4 kJ/mol [19] 27.21 kJ/mol [24] | 21.9±0.2 kJ/mol [19] 21.8 kJ/mol [25] 25.2 kJ/mol [26] |

## 2. Radial thermal expansion of $H_2$-saturated SWNT bundles.

### 2.1. Experimental technique

The radial thermal expansion of $H_2$-saturated bundles of single-walled carbon nanotubes was investigated using a low-temperature capacitance dilatometer (its design and measurement technique are described in [27]). The sample for measurements was prepared by compressing (at 1.1 GPa) the SWNT powder that was used in our investigation of $H_2$ sorption (see previous section). The preparation technique is described elsewhere [3]. According to [28], the applied pressure aligns the CNT axes in a plane normal to pressure vector, the average deviation from the plane being ~4°. The compressed sample was a cylinder 7.2 mm high and 10 mm in diameter, also used in our previous investigations of the radial thermal expansion of pure and Xe-saturated SWNT bundles [2,3].

Prior to measurement, the cell with the sample of pressure-oriented CNTs was evacuated at room temperature for 96 hours to remove possibly available gas impurities. Then it was filled with hydrogen up to 760 Torr and kept at this pressure for 72 hours to saturate the sample with $H_2$. Thereafter, the measuring cell of the dilatometer with the sample in the hydrogen atmosphere was cooled to 12 K and the saturation of the sample was continued. The $H_2$ gas was added to the measuring cell in small portions until the pressure 0.1 Torr was reached. When saturation was completed, the measuring cell was cooled to liquid helium temperature. The thermal expansion was measured in vacuum down to $1 \cdot 10^{-5}$ Torr.

Care was taken to exclude, within the experimental error, the effect of desorption on the LTEC values (see below).

### 2.2. Results and discussion

The temperature dependence of the linear thermal expansion coefficient (LTEC) taken on a $H_2$—SWNT sample in the interval 2.2—27 K is shown in Fig. 2 (curve 1).

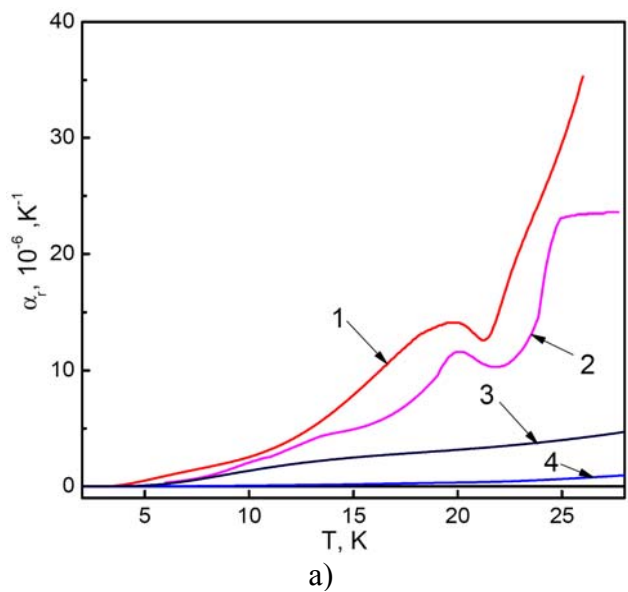

a)

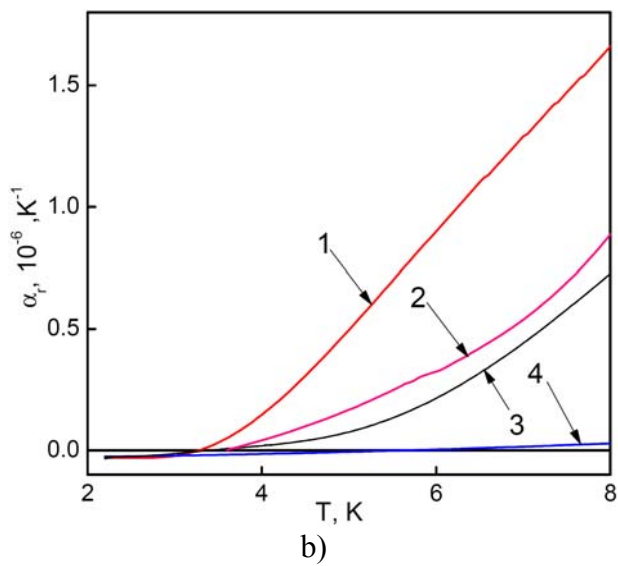

b)

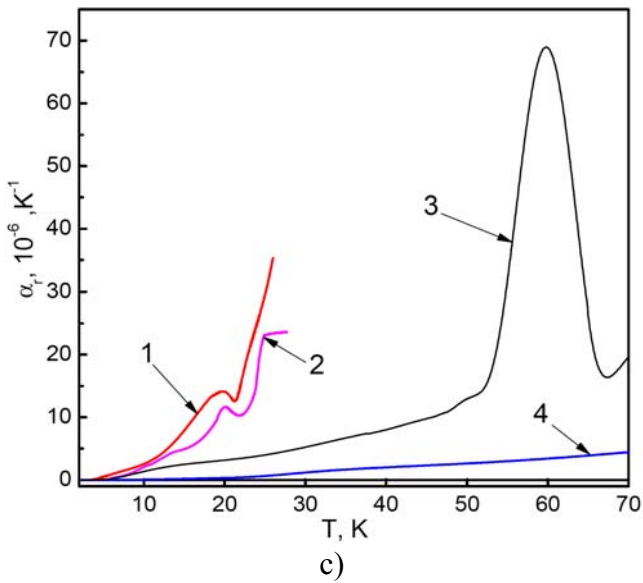

Fig. 2. The radial thermal expansion coefficient of SWNT bundles: 1) $H_2$-saturated; 2) after a partial $H_2$ removal at T = 28 K; 3) Xe-saturated [3]; 4) pure CNTs; a) T = 2.2—28 K; b) T = 2.2—8; c) T = 2.2—70 K.

It is of interest to note the presence of a region with a negative LTEC in the temperature dependence of the thermal expansion of the $H_2$-doped SWNT bundles at T = 2.2—3 K (see Fig. 2b). According to theoretical conclusions [4], the Grüneisen coefficient and the radial thermal expansion of CNTs are negative at rather low temperatures, which is mainly determined by the contribution of the transverse acoustic vibrations perpendicular to the CNT surface. However, the theory [4] was concerned with individual CNTs. In the case of SWNT bundles additional factors contributing to the thermal expansion come into play. Firstly, there appears a positive contribution $\alpha_g$ caused by vibrations of the intertube gaps with temperature. Secondly, the nanotube interaction in the bundles suppresses the negative contribution of the transverse acoustic vibrations perpendicular to the nanotube surfaces [4]. These two positive contributions to the thermal expansion of SWNT bundles decrease both the magnitude and the temperature region of the total negative thermal expansion. The narrowed temperature intervals of negative LTECs in Xe- [3] and $H_2$-doped CNTs manifests the effect of the impurity particles upon the transverse vibrations of the CNTs in the direction perpendicular to their lateral surface. The gases sorbed by SWNT bundles enhance the three-dimensional features of the system, which in turn intensifies its dynamics and increases the positive contribution to the thermal expansion.

It is interesting that the LTECs of the $H_2$—SWNT sample increase sharply at T > 3.5 K (see Fig. 2, curve 1 ($H_2$) and curve 3 (Xe)). This can be due to at least two factors — the molar concentration of $H_2$ (much higher than that of Xe) and the capability of relatively small $H_2$ molecules to reside in SWNT bundles at the sites prohibited for Xe atoms. The impurity that occupies these sites enhances the three-dimensional features of the dynamics of the system and increases the positive contribution to the thermal expansion [21,29,30]. The hydrogen that is inside the nanotubes also serves to increase the LTECs higher than those of the Xe-SWNT samples.

The reproducibility of the results was checked continuously in the course of measurement to be certain that the thermal expansion was unaffected by adsorption.

The thermal expansion was measured through a cyclic procedure: heating to T = $T_0$ + $\Delta T$ ($\Delta T$ = 0.3 ... 1 K), cooling by $\Delta T$ and heating again. The coincidence of the results within the experimental error was taken as an indication of the equilibrium state at $T_0$ (see Fig. 2, curve 1). The disturbance of data reproducibility on cycling signaled that the $H_2$ desorption grew enough to affect the thermal expansion at this temperature. In this case the heating of the sample and the investigation of the temperature dependence of the LTEC were stopped. The LTEC values were reproducible in the interval 2.2—27 K (Fig. 2, curve 1). A significant change in the LTECs in the course of cycling was observed only after heating to T = 28 K (Fig. 1, the temperature region of growing desorption in the diagram) at which the removal of the $H_2$ impurity from the interstitial cavities and grooves of the SWNT bundles becomes intense.

It is reasonable to conclude that the $H_2$ desorption from these sites produces a significant effect on the LTEC magnitudes in the course of measurement. Below T = 28 K the $H_2$ desorption from the areas with low binding energies at the SWNT bundle surface (see Table 2) has no effect, within the experimental error, on the temperature dependence of the radial thermal expansion.

The LTEC measurements were repeated on nanotubes after a partial $H_2$ desorption. For this purpose the sample was held at T = 28 K for an hour and was then cooled to the lowest temperature 2.2 K. After the partial $H_2$ desorption the LTEC results were again reproducible in the interval 2.2—28 K (Fig. 2, curve 2).

The temperature dependence of the radial LTECs of $H_2$-saturated SWNT bundles had maxima at T ≈ 20 K. They were much smaller than in the temperature dependence of the radial TEC of a Xe-doped SWNT sample (see Fig. 2c). Most likely this is because the local maximum in the temperature dependence of the heat capacity and the radial thermal expansion of Xe-saturated SWNT bundles was caused by a change in the spatial configuration due to the redistribution of the impurity molecules. In the case of $H_2$-saturated SWNT bundles such redistribution is less probable because the high concentration of hydrogen leaves far fewer vacant positions for the impurity molecules inside SWNT bundles.

It is possible to assume that the observed maximum is due to the spatial redistribution of the $H_2$ molecules inside the nanotubes, which process is thought to be similar to the melting of solid $H_2$.

## Conclusions

The coefficient of the radial thermal expansion of SWNT bundles saturated with hydrogen has been measured in the interval 2.2—27 K by the dilatometric method. The saturation of SWNT bundles with the normal $H_2$ impurity brings about the following features of the thermal expansions of the $H_2$—SWNT bundle system

1. The temperature interval of the negative thermal expansion of the $H_2$—SWNT bundle system shifts towards lower temperatures as compared to pure SWNT bundles. The negative thermal expansion is caused by the transverse acoustic vibrations of the basic graphene elements of nanotubes. The deposition of impurity molecules on the surface and inside CNTs enhances the three-dimensional features of the system and increases the positive component of the thermal expansion such that the negative component is dominant only at lower temperatures.

2. In the temperature interval 3.5—27 K the radial TEC values of the $H_2$—SWNT system exceed the TECs of pure and Xe-doped nanotubes. The effect is attributed to the influence of at least two factors — the considerably higher molar concentration of $H_2$ in comparison to Xe and the relatively small size of the $H_2$ molecules, making them capable of penetrating into areas in SWNT bundles inaccessible for Xe atoms. This enhances the three-dimensional features of the dynamics of the system and increases the positive contribution to its thermal expansion. Being considerably smaller than Xe atoms, $H_2$ molecules can penetrate in the interstitial channels of SWNT bundles and, through the available surface defects, inside the nanotubes. The hydrogen that is in the interstitial channels loosens the bundles, suppresses the nanotube interaction in the bundles and, finally, leads to an increase in the radial thermal expansion of the system. The $H_2$ molecules that are inside the nanotubes form layers [31] which enhance the three-dimensional features of the system and make a positive contribution to its thermal expansion.

3. The temperature dependences of the radial TEC of the SWNT bundles saturated with hydrogen have weak maxima (T $\approx$ 20 K) that are accounted for by the change in the spatial configuration due to redistribution of the impurity molecules.

The total quantity of sorbed $H_2$ did not increase after the oxidative treatment. Moreover, it was found to decrease slightly. Note that the quantity of sorbed Xe increased considerably after the same treatment of Xe—SWNT samples [3]. The correlation between the sorptive capacities of the starting and oxidated SWNT powders is determined by competition of two mechanisms:

1) The nanotubes of the starting powder had surface defects before the oxidative treatment was applied. The defects let the small $H_2$ molecules pass inside the nanotubes through the interstitial channels and the defects in the CNT walls. The $H_2$ molecules also fills the IC channels of the bundle. These ways are prohibited for large Xe atoms. The oxidates produced by the oxidative treatment in the defect regions could partially block the interstitial channels. This mechanisms suppresses the sorptive capacity of the sample after the oxidative treatment.

2) The oxidative treatment causes a partial disintegration of the SWNT bundles [22]. They decrease in size but grow in number, which extends the total outer surface of the bundles. This mechanism enhances the sorptive capacity of the sample after the oxidative treatment.